\documentclass[conference,a4paper]{IEEEtran}
\usepackage{cite,multirow,graphicx}
\usepackage{morefloats}
\usepackage{soul}
\usepackage{amsmath,bbm}
\usepackage{amsfonts,amssymb,graphicx,subfigure,tikz}
\usepackage{etoolbox}
\usepackage{color}
\usepackage{epstopdf}
\usepackage{enumerate}
\usepackage{siunitx}
\usepackage{mathtools}
\usepackage{hyperref}
\usepackage[ruled,vlined]{algorithm2e}
\newcommand{\nn}{\nonumber}

\begin{document}             
	\title{{On the Optimal Beamwidth of UAV-Assisted Networks Operating at Millimeter Waves}
	}
	\author{\IEEEauthorblockN{Manishika Rawat\IEEEauthorrefmark{1}, Marco Giordani\IEEEauthorrefmark{1}, Brejesh Lall\IEEEauthorrefmark{2}, Abdelaali Chaoub\IEEEauthorrefmark{3}, and Michele Zorzi\IEEEauthorrefmark{1}}
		\IEEEauthorblockA{\IEEEauthorrefmark{1}Department of Information Engineering, University of Padova, Padova, Italy\\
			\IEEEauthorrefmark{2}Department of Electrical Engineering, Indian Institute of Technology Delhi, New Delhi, India\\
			\IEEEauthorrefmark{3}National Institute of Posts and Telecommunications (INPT), Rabat, Morocco
		}\vspace{-1em}\\
		E-mails: manishika.rawat8@gmail.com, giordani@dei.unipd.it, brejesh@ee.iitd.ac.in, chaoub.abdelaali@gmail.com, \\ zorzi@dei.unipd.it
	}
	\vspace{-1em}
	\maketitle
	\begin{abstract}
		The millimeter-wave (mm-wave) bands enable very large antenna arrays that can generate narrow beams for beamforming and spatial multiplexing. However, directionality introduces beam misalignment and leads to reduced energy efficiency. Thus, employing the narrowest possible beam in a cell may not necessarily imply maximum coverage. The objective of this work is to determine the optimal sector beamwidth for a cellular architecture served by an unmanned aerial vehicle (UAV) acting as a base station (BS). The users in a cell are assumed to be distributed according to a Poisson Point Process (PPP) with a given user density. We consider hybrid beamforming at the UAV, such that multiple concurrent beams serve all the sectors simultaneously. An optimization problem is formulated to maximize the sum rate over a given area while limiting the total power available to each sector. 
		We observe that, for a given transmit power, the optimal sector beamwidth increases as the user density in a cell decreases, and varies based on the height of the UAV. Thus, we provide guidelines towards the optimal beamforming configurations for users in rural areas.
		
	\end{abstract}


\begin{tikzpicture}[remember picture,overlay]
	\node[anchor=north,yshift=-20pt] at (current page.north) {\parbox{\dimexpr\textwidth-\fboxsep-\fboxrule\relax}{
			\centering\footnotesize This paper has been accepted for publication at the IEEE Wireless Communications and Networking Conference Workshops (WCNC WKSHPS), 2023. Copyright may change without notice.}};
\end{tikzpicture}	
	\section{Introduction}
	According to the World Bank, 43 \% of the world population lives in rural areas \cite{WBdata}. However, these regions remain mostly unserved while the world prepares to roll out the fifth generation (5G) of mobile networks. According to a report published by the International Telecommunication Union in 2021, the share of Internet users in urban areas is twice the number in rural areas \cite{2021ITUdigital}. The primary cause behind this is the lack of communication infrastructure in remote areas. The next generation of wireless networks (6G) emerges as a solution to this challenge \cite{giordani2020toward}: for example,
	6G is focusing on non-terrestrial networks (NTN) using Unmanned Aerial Vehicles (UAVs), High Altitude Platforms (HAPs), and satellites to promote ubiquitous and high-capacity global connectivity~\cite{giordani2020non}. Notably, these modules can serve as aerial base stations (BS) or to assist the terrestrial BS in providing on-demand, cost-effective coverage in unserved and poorly served areas \cite{Chaoub20216g}.
	
	UAVs, in particular, have been proposed to bridge the digital divide and provide on-demand networks for applications such as disaster management, medical camps, network exploration, and surveillance \cite{wang2020potential, Rahman2022}. 
	Deploying a UAV as a BS (UAV-BS) is quick and affordable compared to a terrestrial network infrastructure, and when operating in the millimeter-wave (mm-wave) bands can promote cost-effective ubiquitous coverage, high throughput, and low latency even in rural areas \cite{matracia2021}.
	
	However, how to deploy UAV-BSs is a challenging design issue, and has been discussed in detail in \cite{elnabty2022survey,alzenad20173}. At the same time, allocating the (limited) resources to the ground users is critical. Several resource allocation problems have been formalized in the literature to meet different quality of service (QoS) requirements such as coverage, fairness, and energy efficiency \cite{zeng2017,xiao2019unmanned, duan2019resource, rupasinghe2018non, yi2019unified, kumar2021dynamic, kaleem2022}. 
	\begin{figure*}[t]
		\begin{center}
			\includegraphics[height=2.1in,width=1.5\columnwidth]{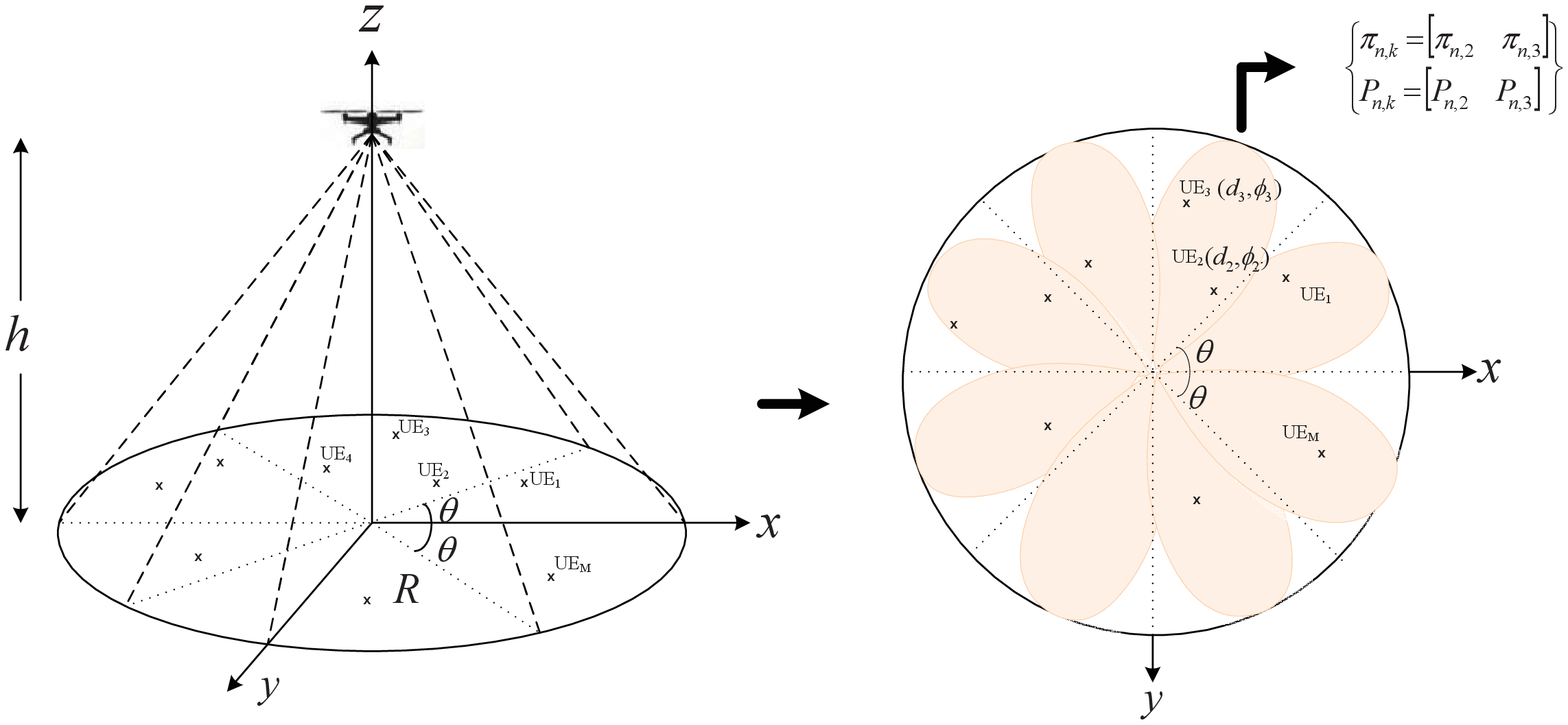}
			\caption{Geometry of the scenario where a UAV-BS serves a cellular area of radius $R$.}
			\label{geometry}
		\end{center}
	\end{figure*}
	An energy-efficient UAV communication system was proposed in \cite{zeng2017} by optimizing the UAV trajectory and jointly considering the communication throughput and the energy consumption. In \cite{xiao2019unmanned}, the authors introduced an approximate beam pattern and provided a solution for the UAV deployment and beam gain allocation to maximize the capacity over a given area. In \cite{duan2019resource}, multi-UAV communication and non-orthogonal multiple access (NOMA) have been combined for the purpose of constructing high-capacity Internet of Things (IoT) uplink transmission systems. The channel assignment, the uplink transmit power of IoT nodes, and the flying heights of UAVs have been jointly optimized to maximize the system capacity. In \cite{rupasinghe2018non}, NOMA transmission with UAV-BS is implemented to provide coverage over a dense user region. A beam scanning approach has been proposed to identify the optimal area within the user region, and hence maximize the achievable sum rate. In \cite{yi2019unified}, the locations of transceivers in the downlink and uplink were modeled using a Poisson Point Process (PPP) or Poisson Cluster Process to derive closed-form expressions of the coverage probability. In \cite{kumar2021dynamic}, an optimal resource allocation problem has been investigated for downlink coverage. It considered concurrent transmission to all sectors, and an asymptotically-optimal solution has been proposed to solve a mixed integer non-linear programming problem. The authors in \cite{kaleem2022} have proposed an intelligent UAV-BS placement and power allocation algorithm to maximize the sum rate in a region. 
	
	The optimal beamwidth for UAV-assisted multi-user systems has been studied in \cite{he2017joint} considering the main lobe of the directional antenna serving a cell. The ground terminals were partitioned into disjoint clusters, which were sequentially served by the UAV, and the joint UAV height and beamwidth have been investigated. The altitude, beamwidth, and location of the UAV and the bandwidth allocated to each user were jointly optimized in \cite{yang2018joint} for uplink UAV communication to minimize the sum uplink power. An algorithm was proposed to obtain a suboptimal solution by assigning different bandwidths to ground terminals. In \cite{tang2019joint}, the UAV location and antenna beamwidth were jointly optimized for quasi-stationary and mobile UAVs. The impact of the beamwidth on UAV location and trajectory has been investigated for minimizing serving time and increasing throughput. All of these works, however, assume that the vertical main lobe of the UAV antenna covers all users. 
	
	In this paper we determine the optimal sector beamwidth for a UAV-assisted cellular network implementing hybrid beamforming, where users are distributed according to a PPP with a given user density. An optimization problem is developed for efficient resource allocation to maximize the sum rate in a sector while ensuring fairness. 
	We observe that the optimal sector beamwidth depends on the user density, the height of the UAV-BS, and the propagation scenario. 
	For a sub-urban scenario, with cell radius of $100$ m and a UAV-BS deployed at a height of $100$ m, the optimal beamwidth decreases from $10^\circ$ to $5^\circ$ as the user density increases from $0.0005$ to $0.002$ UEs/\SI{}{\metre\squared}. On the other hand, for a fixed user density of $0.0005$ UEs/\SI{}{\metre\squared}, the optimal beamwidth initially decreases from $12^\circ$ to $10^\circ$ and then increases to $15^\circ$ as the UAV height increases from $10$ m to $200$ m. 
	We further observe that the number of sectors required to optimally serve a given number of users in an urban region is much larger than that in a rural region for the same QoS requirements.

	
	
	The paper has been organized as follows. In Sec. \ref{system_model} we present our system model; in Sec. III we describe the optimization problem to maximize the sum rate in a cell within given constraints, and present the algorithm we used to compute the sum rate as a function of the sector beamwidth; in Sec. IV we show our simulation results; conclusions are given in Sec. V.
	
	\textbf{Notations:}  For a random variable $X$, $X\sim \text{Poisson}(x)$ denotes that $X$ is Poisson distributed with rate parameter $x$, $X\sim \text{U}(x_1,x_2)$ indicates that $X$ is uniformly distributed in the range $(x_1, x_2]$, and $X\sim \text{Tr}(x_1, x_2)$ denotes that $X$ has triangular distribution in the range $(x_1, x_2]$ \cite{rawat2021statistical}.

	\section{System Model}\label{system_model}
	The system model involves a UAV-BS deployed at a height $h$ and serving a cellular area of radius $R$, as shown in Fig. \ref{geometry}. The cell is divided into $S$ sectors, each of beamwidth $\theta$. The UAV is mounted with a uniform planar array (UPA) antenna of $N$ elements such that $\theta\approx 2/N$ radians \cite{balanis}. The users, represented by crosses in Fig. \ref{geometry}, are distributed in the cell according to a PPP with density $\lambda$. The location of the $k^{th}$ user in a cell is given by $(d_k, \phi_k)$, where $d_k$ is the horizontal distance between the $k^{th}$ user and the UAV-BS and $\phi_k$ is the phase of the $k^{th}$ user measured counterclockwise. Here, $\phi\sim \text{U}(0, 2\pi)$ and $d\sim \text{Tr}(0,R)$. 
	The average number of users in a cell is $M=\pi R^2\lambda$.
	We assume fixed transmit power $P_t$ at the UAV, and orthogonal frequency division multiplexing (OFDM) to serve multiple users on a single beam. Therefore, the total bandwidth $B$ is split into $N_c$ subcarriers to serve multiple users in a sector. We operate at mm-wave frequency in an effort to maximize the communication capacity, and assume hybrid beamforming at the UAV such that all sectors are served by concurrent~beams~\cite{wang2022beamforming}.
	

	The data rate $r_{n,k}$ for the $n^{th}$ subcarrier allocated to the $k^{th}$ user is given by
	\begin{align}
		r_{n,k}=(B/N_c)\log_2(1+P_{n,k}\gamma_{n,k}),	
	\end{align}
	where $\gamma_{n,k}$ is given by \cite{varshney2020optimum}, i.e.,
	\small
	\begin{align}\nonumber
		&\gamma_{n,k}(l_k)=\\
		&\frac{|\mathbbm{h}_{LoS}^{n,k}|^2 PL_{LoS}^{n,k}(l_k)P_r(l_k)+|\mathbbm{h}_{NLoS}^{n,k}|^2 PL_{NLoS}^{n,k}(l_k)(1-P_r(l_k))}{N_0 B/N_cGG_r}.
		\label{gamma_nk}
	\end{align}\normalsize
	In Eq. \eqref{gamma_nk}, $l_k=\sqrt{h^2+d_k^2}$ is the distance between the UAV and the $k^{th}$ user, and $\mathbbm{h}_{LoS}^{n,k}$ and $\mathbbm{h}_{NLoS}^{n,k}$ are the path gains for the line-of-sight (LoS) and non-line-of-sight (NLoS) paths between the UAV and the $k^{th}$ user. Specifically, $k\in\{1, 2, \dots, K_s\}$, where $K_s$ represents the number of users in the $s^{th}$ sector, $s\in\{1, 2, \dots, S\}$, and $n\in\{1, 2, \dots, N_c\}$. $N_0$ is the noise power spectral density, and $G$ and $G_r$ represent the beamforming gains of the transmitting and receiving antennas, respectively. $P_r(l_k)$ is the LoS probability between the UAV and the $k^{th}$ user, and is given~by
	\begin{align}
		P_r(l_k)=\frac{1}{1+e^{\alpha_1\psi^3+\alpha_2\psi^2+\alpha_3\psi+\alpha_4}},
		\label{LoS_pr}	
	\end{align}
	where $\psi=\sin^{-1}\left(h/l_k\right)$ is the elevation angle of the UAV with respect to the $k^{th}$ user, and $\alpha_1$, $\alpha_2$, $\alpha_3$, and $\alpha_4$ are parameters of the LoS probability distribution defined for sub-urban, urban, dense-urban, and high-rise-urban environments in \cite[Table 2]{mohammed2021line}. 
	$PL_{LoS}$ and $PL_{NLoS}$ represent the path losses for LoS and NLoS links; for mm-wave communication~\cite{akdeniz2014millimeter}, they are expressed in dB as
	\begin{align}
		&PL_{LoS}(l_k)=61.4+20\log_{10}(l_k)+\mathcal{N}(0,33.64);\\
		&PL_{NLoS}(l_k)=72.0+29.2\log_{10}(l_k)+\mathcal{N}(0,75.69).
	\end{align}
	where, $\mathcal{N}(\mu,\sigma^2)$ represents a normal random variable with mean $\mu$ and variance $\sigma^2$. In the following section we will develop an optimization problem to maximize the sum rate in a cell while limiting the total power available in a sector.

	\section{Sum Rate Maximization Problem}
	The sum rate maximization problem for a sector can be expressed as:
	\begin{subequations}
		\allowdisplaybreaks
		\begin{alignat}{2}
			&\!\max_{\pi_{n,k},P_{n,k} }        &\quad&       	\sum_{k=1}^{K_s}\sum_{n=1}^{N_c}r_{n,k} \pi_{n,k}\label{eq:OF1}\\
			&\text{s.t.} &  C_{1}:    & \sum_{k=1}^{K_s}\pi_{n,k}\leq 1\quad\forall n,\label{c11}\\
			&                  &  C_{2}:    & \sum_{n=1}^{N_c}r_{n,k}\pi_{n,k}\geq R_0\quad\forall k,\label{c12}\\
			&                  &  C_{3}:    & \sum_{k=1}^{K_s}\sum_{n=1}^{N_c}P_{n,k}\leq \frac{P_t\theta}{360},\label{c13}\\
			&                  &  C_{4}:    & \pi_{n,k}\in \{0,1\}\quad\forall n,k,\label{c14}\\
			&                  &  C_{5}:    & P_{n,k}\geq 0\quad\forall n,k.\label{c15}
		\end{alignat}
	\end{subequations}
	In \eqref{eq:OF1}, we introduce a binary variable $\pi_{n,k}$ to ensure that at least one subcarrier is allocated to one user in a sector in the beam serving time. Here, $\theta$ is the beamwidth of a sector, and $R_0$ is the bit rate. $P_{n,k}$ is the power transmitted over the $n^{th}$ subcarrier allocated to the $k^{th}$ user. Constraints $C_1$ and $C_4$ ensure that the same subcarrier is not allocated to different users, and thus $\pi_{n,k}$ can take only binary values such that the sum across the users is less than or equal to $1$. $C_2$ ensures a minimum data rate to each user. 
	As the cell is divided into sectors, the total power available to a sector would be proportional to the sector beamwidth. Therefore,  $C_3$ limits the total power available to a sector. This limits the maximum data rate available to users as the sector beamwidth decreases or the number of sectors in a cell increases. $C_5$ ensures that the power allocated to each user is non-negative. 
	
	The optimization problem in \eqref{eq:OF1} is a mixed integer non-linear programming problem (MINLP). Due to the non-convexity, the global optimal solution cannot be achieved. The time taken to run this module increases proportionally with the number of users and sectors. However, we can simplify the problem using the mixed integer (MI) property of $\pi_{n,k}$. The simplified optimization problem can be expressed~as
	\begin{subequations}
		\allowdisplaybreaks
		\begin{alignat}{2}
			&\!\max_{\pi_{n,k},P_{n,k} }        &\quad& \sum_{k=1}^{K_s}\sum_{n=1}^{N_c}r_{n,k} \label{OF2}\\\nn
			&\text{s.t.} &  C_{1}, C_4, C_5,   \\
			& &  C_{6}:& \sum_{n=1}^{N_c}r_{n,k}\geq R_0\quad\forall k,\label{c22}\\
			& & C_{7}:& \sum_{k=1}^{K_s}\sum_{n=1}^{N_c}P_{n,k}\leq \pi_{n,k}\frac{P_t\theta}{360},\label{c23}\\
			&& C_{8}:& r_{n,k}\leq \pi_{n,k}R_{max}\quad\forall n,k.\label{c26}
		\end{alignat}
		\label{Op2}
	\end{subequations}

	\SetKwComment{Comment}{/*}{}
	\begin{algorithm}[t!]
		\caption{To compute the sum rate per cell as a function of $\theta$ for a given value of $\lambda$}\label{algo1}
		\SetAlgoLined
		initialize $N_c$, $P_t$, $B$, $R_0$, $G_0$, $G_r$, $N_0$, $R$, $h$, $\lambda$, $\theta\in div(360)$,\\
		\For{each value of $\theta$}{$S=360/\theta$ \Comment*[r]{number of sectors}
			\For{$\eta=1, 2, \dots, \mathbb{N}$\Comment*[r]{MC simulations}}{
				$\chi\sim\text{Poisson}(\pi R^2\lambda)$, 
				$\phi\sim \text{U}(0,2\pi)$, $d\sim \text{Tr}(0,R)$\Comment*[r]{user distribution parameters}
				\For{$s=1, 2, \dots, S$}{Based on the value of $\phi_k$ and $\theta$, determine the users that lie in the $s^{th}$ sector\;
					Compute $g_{n,k}(N_c,K_s)$ for each sector\;
					Run optimization module given in  \eqref{Op2} to allocate $N_c$ subcarriers to $K_s$ users\;
					Compute $r_{n,k}(N_c,K_s)$ for each sector\;
					$Op_s(s)=\sum_{n=1}^{N_c}\sum_{k=1}^{K_s} r_{n,k}(N_c,K_s)$ \Comment*[r]{sum rate per sector}
				}
				$Op_\eta(\eta)=\sum_{s=1}^{S} Op_s$ \\
				$Ap_\eta(\eta)=Op_\eta(\eta)/\chi$ 
			}
			$Op=\sum_{\eta=1}^{\mathbb{N}} Op_\eta/\mathbb{N}$\Comment*[r]{sum rate per cell}
			$Ap=\sum_{\eta=1}^{\mathbb{N}}Ap_\eta/\mathbb{N}$\Comment*[r]{average rate}
		}
	\end{algorithm}
	
	We have omitted $\pi_{n,k}$ from \eqref{OF2} and \eqref{c22} to reduce the redundancy in problem formulation. The condition that $r_{n,k}=0$ when $\pi_{n,k}=0$ is enforced by $C_8$. $C_5$ and $C_7$ ensure that $P_{n,k}$ drops to zero when $\pi_{n,k}=0$. Thus, the algorithm needs to run only for non-zero entries in $\pi_{n,k}$, which reduces the problem complexity. $R_{max}$ in $C_8$ is the maximum data rate that can be achieved. It can be used to limit the number of iterations of the solver to save the processing time.

	The resource allocation problem in \eqref{Op2} can be explained by taking the example of Fig. \ref{geometry}. Here, $S=8$ and we set $N_c = 4$ so that both $\pi_{n,k}$ and $P_{n,k}$ for the second sector measured counterclockwise in Fig. \ref{geometry} are of size $4\times 2$. Thus, the four subcarriers in a sector can be allocated to two users. Because of $C_1$, one of the possible solutions would be $\pi_{n,k}=\begin{pmatrix} 1& 0\\0& 1\\0& 1\\0& 1 \end{pmatrix}$. Accordingly, $P_{n,k}=\begin{pmatrix} P_{1,2}& 0\\ 0& P_{2,3}\\ 0& P_{3,3}\\ 0& P_{4,3} \end{pmatrix}$ such that $\sum_{n=1}^{4}\sum_{k=2}^{3}P_{n,k}\leq P_t/8$. 
	
	Algorithm \ref{algo1} specifies the steps to compute the sum rate per cell and average rate per user as a function of $\theta$ for a given user density. 
	First, we generate users distributed as a PPP in a cell of radius $R$ for a given value of $\theta$ and $\lambda$. The users in the same sector are then categorized for analysis. The optimization function maximizes the sum rate per sector, which is added to produce the sum rate per cell. The average rate per user is computed by dividing the sum rate per cell by the number of users. Here, $\mathbb{N}$ represents the number of Monte Carlo (MC) simulations and $div(x)$ represents the divisors of $x$.

	\section{Results and Discussions}
	In this section, we evaluate the impact of user density and UAV height on the optimal sector beamwidth of a cell. We work with $B=1$ GHz, $P_t=10$ W, $N_c=30$, $R_0=1$ Gbps, $R_{max}=50$ Gbps, $f_c=28$ GHz, and $N_0=-174$ dBm/Hz. The channel is assumed to be Rician with a distribution parameter of $8$ dB \cite{zhou2014ricean}. We use the solving constraint integer program (\textit{scip}) in the General Algebraic Modeling System (GAMS), a high-level modeling system for mathematical optimization to solve the MINLP problem in \eqref{eq:OF1}.
	We consider (i) a rural/sub-urban scenario with a lower user density $\lambda=\{0.0005, 0.0008, 0.001, 0.002\}$ UEs/\SI{}{\metre\squared} in a cell of radius $R=100$ m (Sec. \ref{rural}), and (ii) an urban scenario with a higher user density $\lambda=\{0.05, 0.08, 0.1\}$ UEs/\SI{}{\metre\squared} in a cell of radius $R=10$ m (Sec. \ref{urban}). The impact of the UAV height is explored in Sec. \ref{UAV_height}. We assume $G=N$ and $G_r=1$ for the analysis \cite{balanis}. The results have been obtained for $500$ MC simulations.
	\begin{figure}[t!]
		\centering  
		\subfigure[Sum rate per cell.] {\includegraphics[width=\columnwidth,height=2.3in]{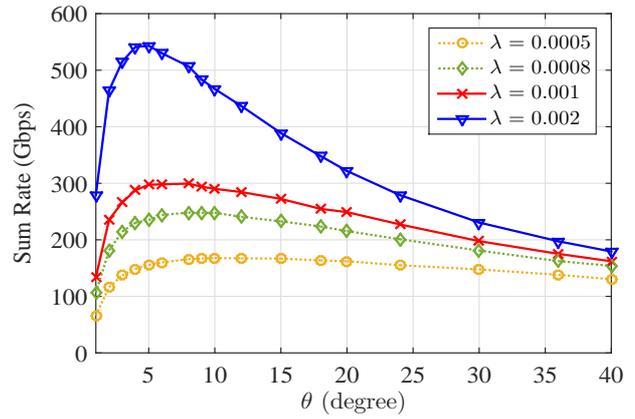}\label{SR_R100}} 
		\subfigure[Average rate per user.] {\includegraphics[width=\columnwidth,height=2.3in]{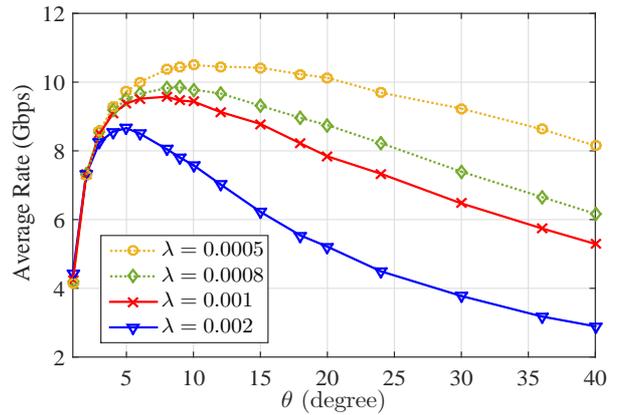}\label{ASR_R100}}
		\subfigure[Jain's fairness index.]
		{\includegraphics[width=\columnwidth,height=2.3in]{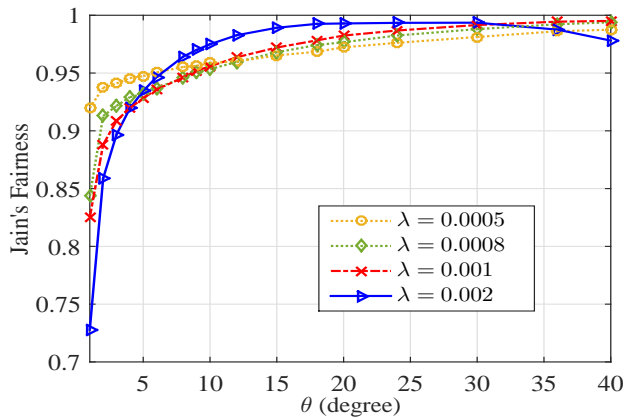}\label{JF_R100}}
		\caption{Sum rate per cell, average rate per user, and Jain's fairness index in a rural/sub-urban scenario with $R=100$ m and $h=100$ m, as a function of $\theta$, for different user densities.}\label{Sum_rate_rural}
	\end{figure}
	\subsection{Rural/Sub-Urban Scenario}\label{rural}
	Fig. \ref{SR_R100} plots the sum rate obtained by solving the optimization problem in \eqref{Op2} as a function of $\theta$. The result is obtained for $R=100$ m, $h=100$ m, and different values of $\lambda$. 
	First, we observe that as the user density increases, the number of users in the cell increases, thus increasing the sum rate per cell, which validates the accuracy of our results.	Initially, as $\theta$ increases the sum rate also increases given that the power allocated to each sector increases proportionally. However, after a certain value of $\theta$, the sum rate decreases exponentially. This can be attributed to the lower spatial reuse as $\theta$ increases, i.e., as the number of sectors decreases. Therefore, we conclude that the smallest beamwidth does not necessarily ensure the maximum sum rate. The sector beamwidth at which the sum rate is maximized is represented by $\theta_{opt}$. According to Fig. \ref{SR_R100}, $\theta_{opt}=10^\circ$ for $\lambda=0.0005$. As $\lambda$ increases from $0.0008$ to $0.002$, $\theta_{opt}$ decreases from $9^\circ$ to $5^\circ$, respectively. This implies that in a rural environment, it is desirable to operate through wider beams as the density of users decreases. 
	%
	%
	
	Fig. \ref{ASR_R100} plots the average rate per user as a function of $\theta$ for different values of $\lambda$. The use of the mm-wave technology ensures that the rate is always above 1 Gbps, and in line with the requirements of most 5G/6G applications. As expected, as the number of users in the cell increases, the average rate per user decreases.
	Moreover, as $\theta$ increases the average rate also increases, and follows the same trend of the sum rate as explained in the previous paragraph. However, $\theta_{opt}$ decreases as the user density increases. This is because, as the number of users in each sector grows, the amount of power allocated to each user decreases. Consequently, in order to maximize the sum rate, the objective function drives $\theta$ towards smaller values. Therefore, we conclude that for a densely populated rural region, a higher number of sectors would be optimal. 
	
	\begin{figure}[t!]
		\centering  
		\subfigure[Sum rate per cell.] {\includegraphics[width=\columnwidth,height=2.3in]{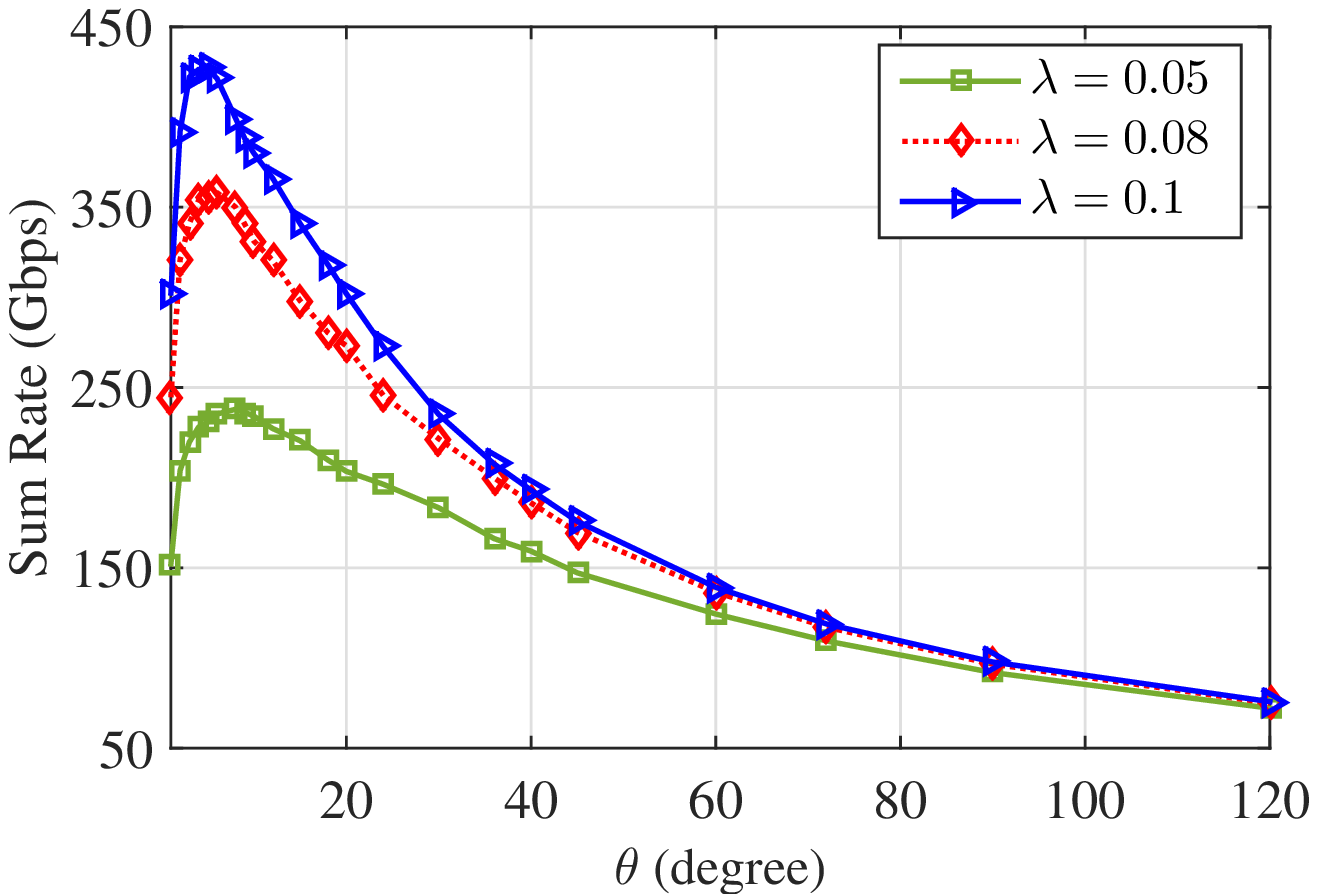}\label{SR_R10}} 
		\subfigure[Average rate per user.] {\includegraphics[width=\columnwidth,height=2.3in]{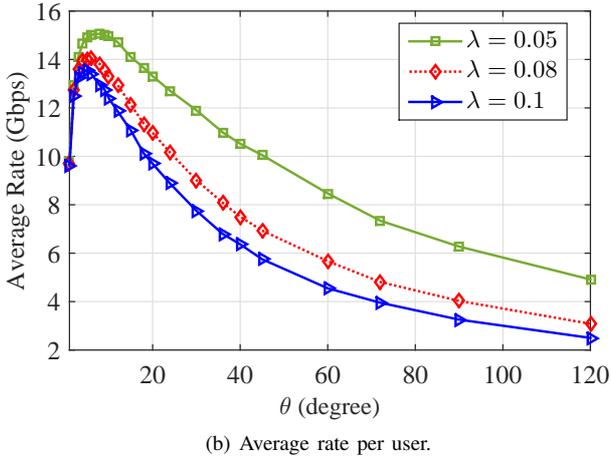}\label{ASR_R10}}
		\caption{Sum rate per cell and average rate per user in an urban scenario with $R=10$ m and $h=20$ m as a function of $\theta$, for different user densities.}\label{SAR_R10}
	\end{figure}
	
	In order to demonstrate the impact of the fairness among the users in a cell, Fig. \ref{JF_R100} plots Jain's Fairness index as a function of $\theta$ for different user densities. It is computed as $J=\dfrac{\left(\sum_i R_i\right)^2}{M\sum_i R_i^2}$, where $R_i$ represents the rate per user \cite{evangelista2019fairness}. The value of Jain's fairness index for $\lambda=\{0.0005, 0.0008, 0.001, 0.002\}$ at $\theta_{opt}$ is $\{0.96, 0.9504, 0.9461, 0.9345\}$, respectively. We observe that $J$ increases with $\theta$, and good fairness is achieved at the optimal sector beamwidth. This can be attributed to constraint $C_6$ in \eqref{Op2}, where a minimum data rate is guaranteed to each user. Thus, we conclude that the proposed resource allocation ensures fairness among all the users in a cell.

	\subsection{Urban Scenario}\label{urban}
	Fig. \ref{SR_R10} plots the sum rate as a function of $\theta$ for various user densities in a cell of radius $R=10$ m. The UAV is deployed at $h=20$ m. As observed in Sec.~\ref{rural}, the sum rate initially increases and then decreases exponentially after $\theta$ reaches its optimal value. However, $\theta_{opt}$ obtained here is generally smaller than in the rural/sub-urban scenario. At $\lambda=0.05$, $\theta_{opt}=8^\circ$ which gives $M=15.7$ and $S=45$ to serve the users optimally. As the user density increases from $\lambda=0.08$ to $0.1$, $\theta_{opt}$ decreases from $6^\circ$ to $5^\circ$. This implies that the number of sectors required to serve all the users increase from $S=60$ to $S=72$, respectively. 
	A similar trend for the value of $\theta_{opt}$ is observed in the plot of the average rate per user in Fig. \ref{ASR_R10}. 
	Notice that these results were obtained up to a beamwidth of $120^\circ$, with the cell having only three sectors to provide a more comprehensive picture.

	%
	
	\begin{figure}
		\begin{center}
			\includegraphics[height=2.2in,width=\columnwidth]{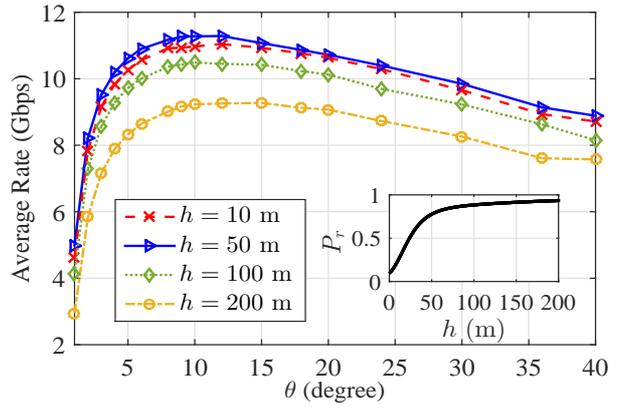}
			\caption{Average rate per user vs. $h$ and $\theta$ with $R=100$ m and $\lambda=0.0005$.}
			\label{JF_R10}
		\end{center}
	\end{figure}
	
	\subsection{Impact of the UAV Height}\label{UAV_height}
	
	Fig. \ref{JF_R10} plots the average rate per user as a function of $\theta$ and $h$ in a rural/sub-urban cell with radius $R=100$ m and $\lambda=0.0005$ UEs/\SI{}{\metre\squared}. The LoS probability ($P_r$) as a function of $h$ defined in Eq. \eqref{LoS_pr} is shown in the inset. We observe that $P_r$ is small for $h\leq50$ m. As a result, the average rate per user for $h=50$ m is higher than for $h=10$ m. As $h$ continues to increase, the LoS probability increases, but so does the path loss. The impact of the path loss becomes the dominant factor when $h>100$ m, even though the LoS probability also increases. The effect is also visible in the value of $\theta_{opt}$, which decreases initially from $12^\circ$ to $10^\circ$ as $h$ increases from $50$ m to $100$ m, and then increases to $\theta_{opt}=15^\circ$ at $h=200$ m. This is because, for $50\leq h\leq 100$ m, the path loss is relatively small, thus the optimization problem drives $\theta$ to a smaller value to increase the antenna gain and the sum rate. However, to compensate for the very high path loss at $h>100$ m, the optimization problem attempts to increase the power allocated to the users by increasing $\theta_{opt}$. Therefore, we conclude that the optimal sector beamwidth is a function of~$h$.
	
	In order to compare the optimal number of sectors required to serve a given number of users in the rural/sub-urban and urban scenarios, we consider two configurations with the same LoS probability, i.e., $R=100$ m and $h=50$ m, and $R=10$ m and $h=20$ m, respectively. For the first configuration with $\lambda=0.0005$ and $M=15.7$, the optimal number of sectors at $\theta_{opt}$ is $S=30$.  This is much lower than $S=45$ obtained in the second configuration with the same number of users in Sec. IV-B. Consequently, the number of elements required in a UPA antenna decreases from $N=15$ in an urban environment to around $N=12$ in a sub-urban scenario. This implies a considerable reduction in the cost and complexity of the antenna system for serving a rural/sub-urban area. 
	
	

	\section{Conclusion}
		\vspace{-.3em}
	In this work, we investigated the optimal sector beamwidth for a cell served by a UAV acting as a BS. Assuming that users are distributed according to a PPP, we observed that there is an optimal beamwidth to maximize the sum rate over a region. 
	For a given transmit power, this optimal beamwidth is a function of the user density and the height of the UAV. 
	Based on simulations we observed that, for the same LoS probability, $S=45$ sectors are required in an urban region vs. $S=30$ sectors in a rural region to optimally serve an average number of $15.7$ users in a cell with the same QoS requirements. 
	Also, we showed that a remote area with lower density of users can be optimally served by a UAV-BS flying at lower altitude. This implies lower complexity in terms of antenna size and transmit power, and less spatial reuse and interference among the sectors, respectively, compared to the urban environment. Our conclusions encourage the use of UAV-BSs to connect remote and poorly served areas.
	

	\vspace{-.5em}
	\bibliographystyle{IEEEtran}
	\bibliography{reference_UAV}
	\vspace*{-0.5em}
\end{document}